\documentclass[twocolumn,showpacs,preprintnumbers,amsmath,amssymb]{revtex4}
\usepackage{epsfig}

\usepackage{graphicx}
\usepackage{dcolumn}
\usepackage{bm}

\newcommand{\be}{\begin{equation}}
\newcommand{\ee}{\end{equation}}
\newcommand{\bea}{\begin{eqnarray}}
\newcommand{\eea}{\end{eqnarray}}
\newcommand{\beq}{\begin{equation}}
\newcommand{\eeq}{\end{equation}}
\newcommand{\nn}{\nonumber}

\def\ga{\mathrel{\mathpalette\fun >}}
\def\fun#1#2{\lower3.6pt\vbox{\baselineskip0pt\lineskip.9pt
\ialign{$\mathsurround=0pt#1\hfil##\hfil$\crcr#2\crcr\sim\crcr}}}

\begin{document}

\title{Asymptotic regime for hadron-hadron diffractive collisions at
ultrahigh energies}

\author{V.V. Anisovich}
\affiliation{National Research Centre "Kurchatov Institute",
Petersburg Nuclear Physics Institute, Gatchina 188300, Russia}
\author{V.A. Nikonov}
\affiliation{National Research Centre "Kurchatov Institute",
Petersburg Nuclear Physics Institute, Gatchina 188300, Russia}
\affiliation{Helmholtz-Institut f\"ur Strahlen- und Kernphysik,
Universit\"at Bonn, 53115 Bonn, Germany}
\author{J. Nyiri}
\affiliation{Institute for Particle and Nuclear Physics, Wigner RCP,
Budapest 1121, Hungary}

\date{\today}

\begin{abstract}
Using the pre-LHC and LHC data for $\pi p$ and
$pp$ diffractive collisions we study the ultrahigh energy asymptotic
regime in the framework of the black disk picture. The black disk
picture, being constrained by the $s$-channel unitarity condition
and the $t$-channel analyticity, gives rather definite predictions
for diffractive processes increasing with the energy. To deal with
the data, we consider the Dakhno-Nikonov eikonal model which
predicts a growth of the $\ln^2s$ type for total and elastic cross
sections and $(\tau={\bf q}_\perp^2\ln^2s)$-scaling for diffractive
scattering and diffractive dissociation of hadrons. According to the
calculations, ultrahigh energy asymptotic characteristics of
diffractive and total cross sections are universal, and this results
in the asymptotic equality of cross sections for all types of
hadrons. We estimate the energy scale of the asymptotics in
different processes. The manifestation of the asymptotic regime in
hadron fragmentation reactions is discussed.
\end{abstract}
\maketitle

\section{Introduction}

Experimental data for diffractive high energy collisions
\cite{pre,totem,auger} definitely tell us that only
at $\sqrt{s}\sim 7$ TeV we approach
the energy region which can be called an asymptotic one. These are
energies at which a picture of the black disk appears in the impact
parameter space $\bf b$, see \cite{block,ryskin,ann}. Actually, it
is a black spot at $|\bf{b}|\simeq 0$; according to estimations only
at $\sqrt{s}\sim 100$ TeV can one observe a shaped disk \cite{ann}.
The problem we discuss here concerns possible    predictions in the
asymptotic energy region $\sqrt{s}\ga 100$ TeV and the range of
accuracy of the predictions.

The observed growth of total cross sections at pre-LHC energies
\cite{pre} prompts   investigations of models such as with a maximal
increase allowed by the Froissart bound \cite{Froi} or with
power-$s$ behaviour \cite{Kaid,Land}. Taking into account the
$s$-channel unitarization of the scattering amplitude leads to
damping of  power-$s$ growth to the $(\ln^2 s)$-type, see
\cite{Gaisser,Block,Fletcher}. Still, let us emphasize that
exceeding the Froissart bound does not violate the general
constraints for the scattering amplitude \cite{azimov}. Presently
the problem of unitarization of high energy amplitudes which
correspond to increasing diffractive cross sections is a subject of
intensive studies, see, for example,
\cite{1110.1479,1111.4984,sch-rys,martynov} and references therein.

A model for high-energy
$\pi p$ and  $p^\pm p$ collisions  was suggested by Dakhno and
Nikonov \cite{DN} and successfully used for the description of the
diffractive pre-LHC data, $\sqrt s \sim 0.2-1.8$ TeV.
The model takes into account the quark structure of colliding
hadrons, the gluon origin of the input pomeron and the colour
screening effects in collisions. The model can be considered as a
realization of the Good-Walker eikonal approach \cite{GW} for a
continuous set of channels.

In the paper \cite{ann} the diffractive $pp$-scattering was
considered in terms of the Dakhno-Nikonov model, concentrating the
attention on the ultra-high energy behaviour. The $pp$ data were
re-fitted taking into account new results in the TeV-region
\cite{totem,auger}. The region 5-50 TeV turns out to be the one
where the asymptotic behaviour starts; the asymptotic regime should
reveal itself definitely at $10^2-10^4$ TeV.

For the ultra-high energy limit the Dakhno-Nikonov model predicts
for total and elastic hadron-hadron cross sections a
$(\ln^2s)$-growth: $\sigma_{tot}\sim \ln^2s,\quad
\sigma_{el}\sim\ln^2s$ and
 $[\sigma_{el}/\sigma_{tot}]_{\ln s\to\infty}\to 1/2 $.
The high energy cross sections ($\sigma_{el}$, $\sigma_{tot}$)
approach    their asymptotic values from bottom to top:
$\sigma_{tot}(s)/\sigma_{tot}^{(asym)}(s)< 1 $; this gives the
illusion of exceeding the Froissart bound (see, for example,
discussion in \cite{block3,1212.5096}). Further, the model predicts
that differential elastic cross sections depend asymptotically on
transverse momenta with a relation for $\tau$-scaling:
 $d\sigma_{el}(\tau)/d\tau= D(\tau )$ with
 $\int_0^\infty d\tau D(\tau)=\sigma_{el}(s)$
 and
 $\tau={\bf q}_\perp^2\sigma_{tot}\propto {\bf q}_\perp^2\ln^2s \,$ .

The model points to the universal behaviour of all total (and
elastic) cross sections. It is the consequence of the universality
of     colliding disk structure (or the structure of parton clouds)
at ultrahigh energy. But the question is at what energy range the
asymptotic regimes are switched on for different processes; we have
calculated here $\sigma_{tot}$, $\sigma_{el}$, $\sigma_{inel}$ for
$pp$, $\pi p$ and $\pi \pi$ collisions.

Further, we consider diffractive dissociation processes. We
demonstrate that these processes are
increasing at asymptotic energies
($\sigma_{D}\propto\ln{s}$, $\sigma_{DD}\propto\ln{s}$)
but their relative contribution tends to zero
($\sigma_{D}/\sigma_{tot}\to 0$, $\sigma_{DD}/\sigma_{tot}\to 0$).

\section{ Diffractive scattering cross sections}

The model is based on the hypothesis of     gluonic origin of
$t$-channel forces, and these gluons form pomerons. Hadrons, mesons
(two-quark composite systems) and baryons (three-quark composite
systems) scatter on the pomeron cloud. It is supposed that the
pomeron cloud behaves as a low-density gas, and pomeron-pomeron
interactions, as well as $t$-channel transitions of the type $P\to
PP$, $P\to PPP$ and so on, can be neglected (for details see
\cite{DN}).

The pomerons are formed by effective gluons ($G$) which are massive,
$\sim 700-1000$ MeV \cite{parisi,field}. The pomeron parameter
$\alpha'_P$ is small $\alpha'_P\simeq(0.10-0.25)$ GeV$^{-2}$, which
means that pomerons are relatively heavy and hard
\cite{gribov-hard}. The gluon structure of the pomeron provides
colour screening effects for hadron quarks \cite{adnPR}.

\subsection{Formulae of the eikonal approach}

Below we present formulae of the Dakhno-Nikonov model for the
eikonal cross sections and then briefly discuss the used
parametrization.

The total and elastic cross sections are:
\begin{widetext}
\bea \label{4}
\sigma_{tot}(AB)&=&2\int d^2b \int dr'\varphi^2_A(r')dr''
\varphi^2_B(r'')
\left [1-\exp{(-\frac{1}{2}
\chi_{AB}(r',r'',{\bf b})})\right], \nn \\
\sigma_{el}(AB)&=&\int d^2b
 \bigg(\int dr'\varphi^2_A(r')dr''\varphi^2_B(r'')
\left[1-\exp{(-\frac{1}{2}\chi_{AB}(r',r'',{\bf b})})\right]
\bigg)^2 .
\eea
\end{widetext}
Here $dr\varphi^2_{A}(r)$, $dr\varphi_{B}^2(r)$ are the quark densities
of  colliding hadrons:
\bea \label{5}
dr \varphi^2_p(r)&=&d^2r_1d^2r_2d^2r_3 \delta ^{(2)}( {\bf r}_1+{\bf
r}_2+{\bf r}_3) \varphi^2_p(r_1,r_2,r_3), \nn \\
dr \varphi^2_{\pi}(r)&=&d^2r_1d^2r_2\delta ^{(2)}( {\bf r}_1+{\bf r}_2)
\varphi^2_{\pi}(r_1,r_2),
\eea
 where ${\bf r}_a$ are the transverse coordinates of  quarks, and
$\varphi_{A}^2$, $\varphi_{B}^2$ are given by quark wave functions
squared integrated over longitudinal variables. Proton and pion
quark densities are determined using the corresponding form factors;
such an estimation can be found, for example, in \cite{AMN}.

The profile-function $\chi_{AB}$ describes the interaction of quarks
via pomeron exchange as follows:
\bea
\label{yf5}
\chi_{AB}(r',r'',{\bf b})&=&\int d^2b'd^2b''\delta^{(2)} ({\bf b}-{\bf b'}+{\bf
b''})
\nn \\
&\times&
S_A(r',{\bf b'})S_B(r'',{\bf b''}), \nn  \\
S_{\pi}(r,{\bf b})&=&\rho( {\bf b}- {\bf
r}_1)+\rho( {\bf b}- {\bf r}_2)
\nn \\
&-& 2\rho( {\bf b}-\frac{ {\bf r}_1+
{\bf r}_2}{2}) \exp(-\frac{( {\bf r}_1-
 {\bf r}_2)^2}{4r^2_{cs}}), \nn \\
S_p( {\bf r}, {\bf b})&=&
\sum\limits_{i=1,2,3}\,\rho( {\bf b}- {\bf r}_i)
\nn \\
&-&\sum\limits_{i\ne k}\,
\rho( {\bf b }-\frac{ {\bf r}_i+ {\bf r}_k}{2})
\exp(-\frac{( {\bf r}_i-
 {\bf r}_k)^2}{4r^2_{cs}}).\nn \\
\eea
The term $\rho( {\bf b}- {\bf r}_i)$ describes the diagram
where the pomeron is connected to one of the hadron quarks while the
terms proportional to $\exp(-r^2_{ij}/r^2_{cs})$ are related to the
diagram where the pomeron is connected to two quarks of the hadron.
Such a diagram is a three-reggeon graph $GGP$ where $G$ is the
reggeized gluon. Functions $S_{\pi}$ and $S_p$ tend to zero as
$|{\bf r}_{ij}| \to 0$: this is the colour screening phenomenon
inherent to the gluonic pomeron. For the sake of convenience one can
perform calculations in the centre-of-mass system of the colliding
quarks, supposing that the hadron momentum is shared equally between
its quarks. Then
\bea \label{yf7}
\rho({\bf b})&=& \frac
{g}{4\pi(G+\frac{1}{2}\alpha'_P\ln{\frac{s}{s_0}})}
 \exp\left [-\frac{{\bf b}^2}{4(G+\frac{1}{2}\alpha'_P\ln{\frac{s}{s_0}})}\right],
\nn  \\
g^2&=&g_0^2+g_1^2\left ( \frac{s_{qq}}{s_0} \right )^{\Delta}
\eea
 with the energy squared of the colliding quarks $s_{qq}$ and
$s_0=1$ GeV$^2$. The parametrization of $g^2$ corresponds to the
two-pole form of the QCD-motivated pomeron with intercepts
$\alpha(0)=1$ and $\alpha(0)=1+\Delta$.

As it was mentioned, the Dakhno-Nikonov model is actually a
realization of the Good-Walker eikonal approach \cite{GW} for a
continuous set of channels: each quark configuration with fixed
coordinates is a separate channel. The two-pole pomeron exchange is
popular from the sixties till now, see for example ref. \cite{DL}.

\subsection{Inelastic diffractive cross sections   }

The diffractive cross section for the two-particle reaction
$A_1B_1\to A_2B_2$ in the Dakhno-Nikonov eikonal approach reads:
\begin{widetext}
\begin{eqnarray}
(2\pi)^2\frac{d^2\sigma_{el}}{d^2q_\perp}(A_1B_1\to A_2B_2)&=&
\int d^2b\ e^{i{\bf q_\perp} {\bf b}}
\int d^2\tilde b\ e^{-i{\bf q_\perp}{\bf \tilde b}}
\nn \\ &\times&
\int dr'dr''\varphi_{A_1}(r')
 \varphi_{B_1}(r'')
\left[1-\exp{(-\frac{1}{2}\chi_{AB}(r',r'',{\bf b})})\right]
\varphi_{A_2}(r')\varphi_{B_2}(r'')
\nn \\ &\times&
\int d\tilde r'd\tilde r''\varphi_{A_1}(\tilde r')
 \varphi_{B_1}(\tilde r'')
\left[1-\exp{(-\frac{1}{2}\chi_{AB}(\tilde r', \tilde r'',{\bf\tilde
b})})\right]
\varphi_{A_2}(\tilde r')\varphi_{B_2}(\tilde r'')\,.
\label{yf9}
\end{eqnarray}
\end{widetext}
 As a consequence of the universality of $\rho ({\bf r})$, the block
 $\left[1-\exp{(-\frac{1}{2}\chi_{AB}(r',r'',{\bf b})})\right]$ which
 is responsible for the interaction is universal and depends only on
the type of colliding hadrons, mesons or baryons. This leads to the
universal behaviour of cross sections at ultrahigh energies. The
universality appears at energies when the essential values of $|\bf
b|$ are much larger than the average interquark distances, $|\bf
b|>>r$; in this region the integrations over impact parameters and
interquark distances are in fact separated.

A hadron in diffractive collision (to be definite, let it be $A_1$)
can produce a set of similar states: these are transitions $A_1\to
A_1,\,A_2,\,A_3,\,\ldots$. If the produced hadrons belong to a complete
set of states, $\sum_n A_n\rangle\langle A^+_n =I$, the sum of such
processes gives a cross section $AB\to X_A +B$ which includes
elastic and diffraction dissociation processes $ X_A=A +X_{DA}$. This
sum of cross sections is equal to:
\begin{widetext}
\bea \label{yf10}
(2\pi)^2\frac{d\sigma_{X(A)}}{d^2q_\perp}(AB\to X_A +B )&=& \int
d^2b \int d^2\tilde b \ e^{i{\bf q_\perp} {\bf b}} \ e^{-i{\bf
q_\perp}{\bf \tilde b}} \int dr'dr''d\tilde r'' \varphi^2_A(r')
\varphi^2_B(r'')\varphi^2_B(\tilde r'')
\nn \\
&\times&
\left[1-\exp{(-\frac{1} {2}\chi_{AB}(r',r'',{\bf b})})\right]
\left[1-\exp{(-\frac{1}{2}\chi_{AB}( r',\tilde r'',
{\bf \tilde b})})\right].
\eea
The cross section integrated over momenta transfer reads
\bea \label{yf11}
\sigma_{X(A)}(AB\to X_A +B )&=&
\int d^2b
\int dr'dr''d\tilde r''
\varphi^2_A(r')
\varphi^2_B(r'')\varphi^2_B(\tilde r'')
\nn \\ &\times&
\left[1-\exp{(-\frac{1}{2}\chi_{AB}(r',r'',{\bf b})})\right]
\left[1-\exp{(-\frac{1}{2}\chi_{AB}(r',\tilde r'',
{\bf  b}))}\right].
\eea
 Let us recall that Eq. (\ref{yf11}) gives us a sum of elastic and
diffraction dissociation cross sections $\sigma_{X(A)}(AB\to X_A +B
)= \sigma_{el}(AB)+\sigma_{D(A)}(AB)$.
The sum of elastic and quasi-elastic
cross sections $\sigma_{X(A)X(B)}(AB\to X_A +X_B )$ is
determined by
the relation:
\bea \label{dd}
\sigma_{X(A)X(B)}(AB\to X_A +X_B )&=&
\sigma_{el}(AB)+\sigma_{D(A)}(AB)+\sigma_{D(B)}(AB)+
\sigma_{D(A)D(B)}(AB )=\nn \\
&=&\int d^2b \int dr'dr'' \varphi^2_A(r') \varphi^2_B(r'')
\left[1-\exp{(-\frac{1}{2}\chi_{AB}(r',r'',{\bf b})})\right]^2 .
\eea
\end{widetext}
At $\ln s \to\infty$ this value tends to $\frac12 \sigma_{tot}$
from bottom to top \cite{pumplin}:
$[\sigma_{el}(AB)+\sigma_{D(A)}(AB)+\sigma_{D(B)}(AB)+
\sigma_{D(A)D(B)}(AB )]_{\ln s\to \infty}\to \frac12 \sigma_{tot}$,
approaching the
asymptotic regime  for $\sigma_{X(A)X(B)}(AB\to X_A +X_B )$ is as
quick as for $\sigma_{tot}(AB)$. Besides it means that at $\ln s\to
\infty$ one has $[\sigma_{D(A)}(AB)+\sigma_{D(B)}(AB)+
\sigma_{D(A)D(B)}(AB)]/\sigma_{tot}(DD)\to 0$.

But the experimental specification of the
diffraction dissociation cross sections involved in Eq. (\ref{dd})
faces the problem of separation from events determined by the inner
structure of the colliding disk, for example, that due to the
three-pomeron diagram processes.

\begin{figure}[ht]
\centerline{\epsfig{file=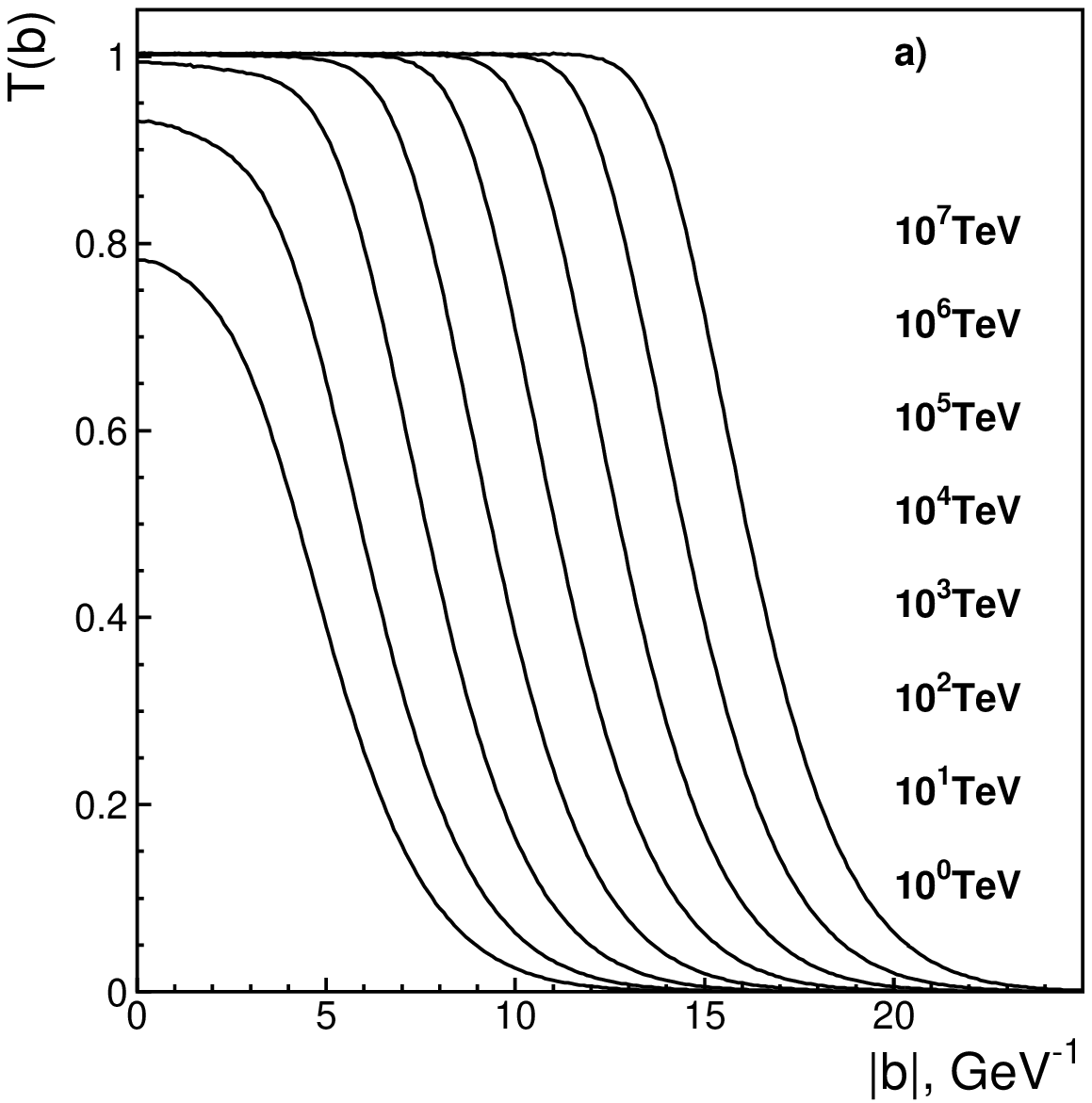,width=8cm}}
\centerline{\epsfig{file=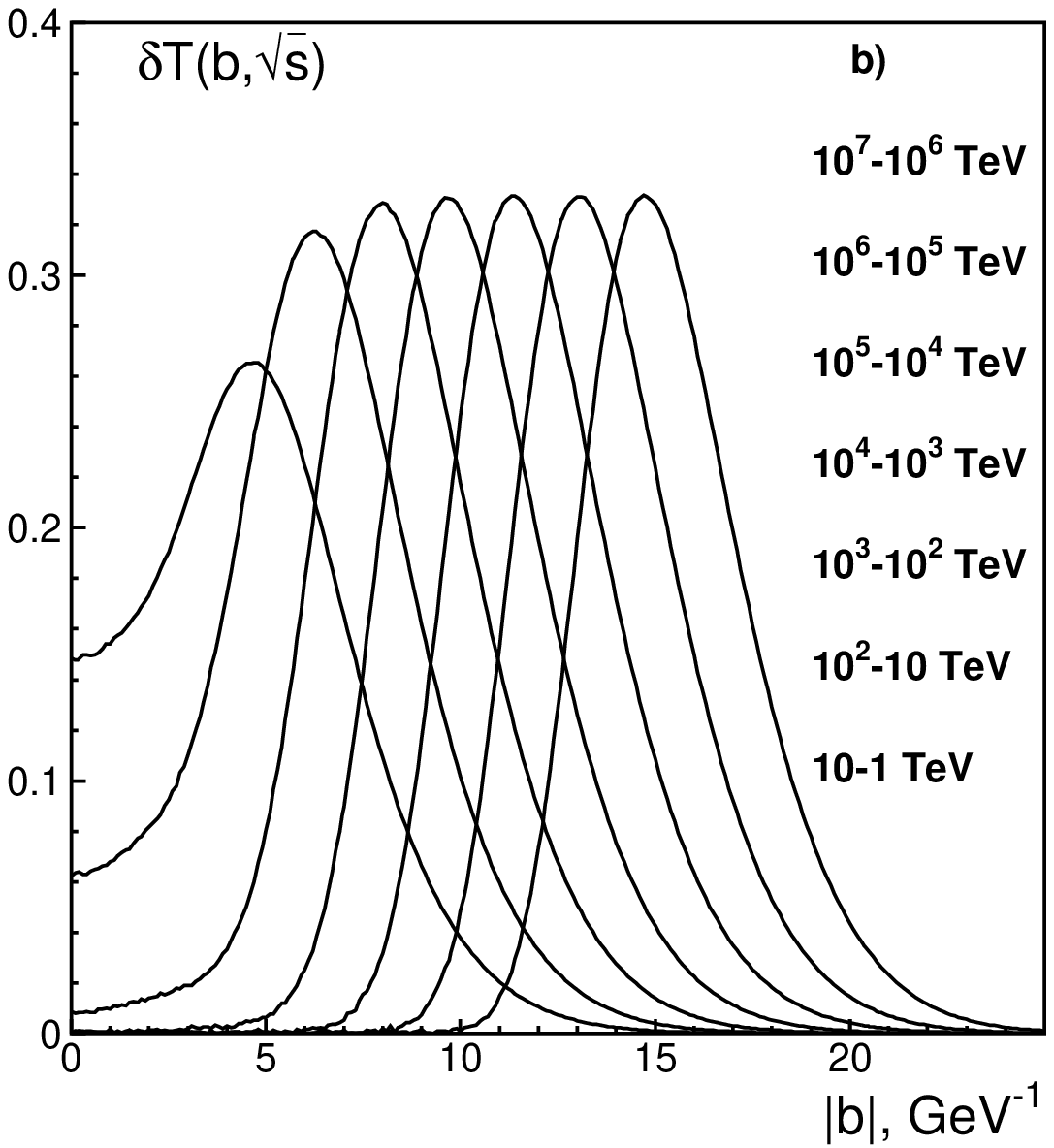,width=8cm}}
\caption{
 a)  Profile functions $T(b)$ determined in Eq. (\ref{yf13})
 at a set of energies $\sqrt{s}=1,10,100,...,10^7$ TeV;
 b) The profile function growth factor determined as
$\delta T(b, \sqrt{s})=[T(b,10 \sqrt{s})-T(b, \sqrt{s})]$
for $\sqrt{s}=1,10,100,...10^6$ TeV.}
\label{f5}
\end{figure}

\begin{figure}[h]
\centerline{\epsfig{file=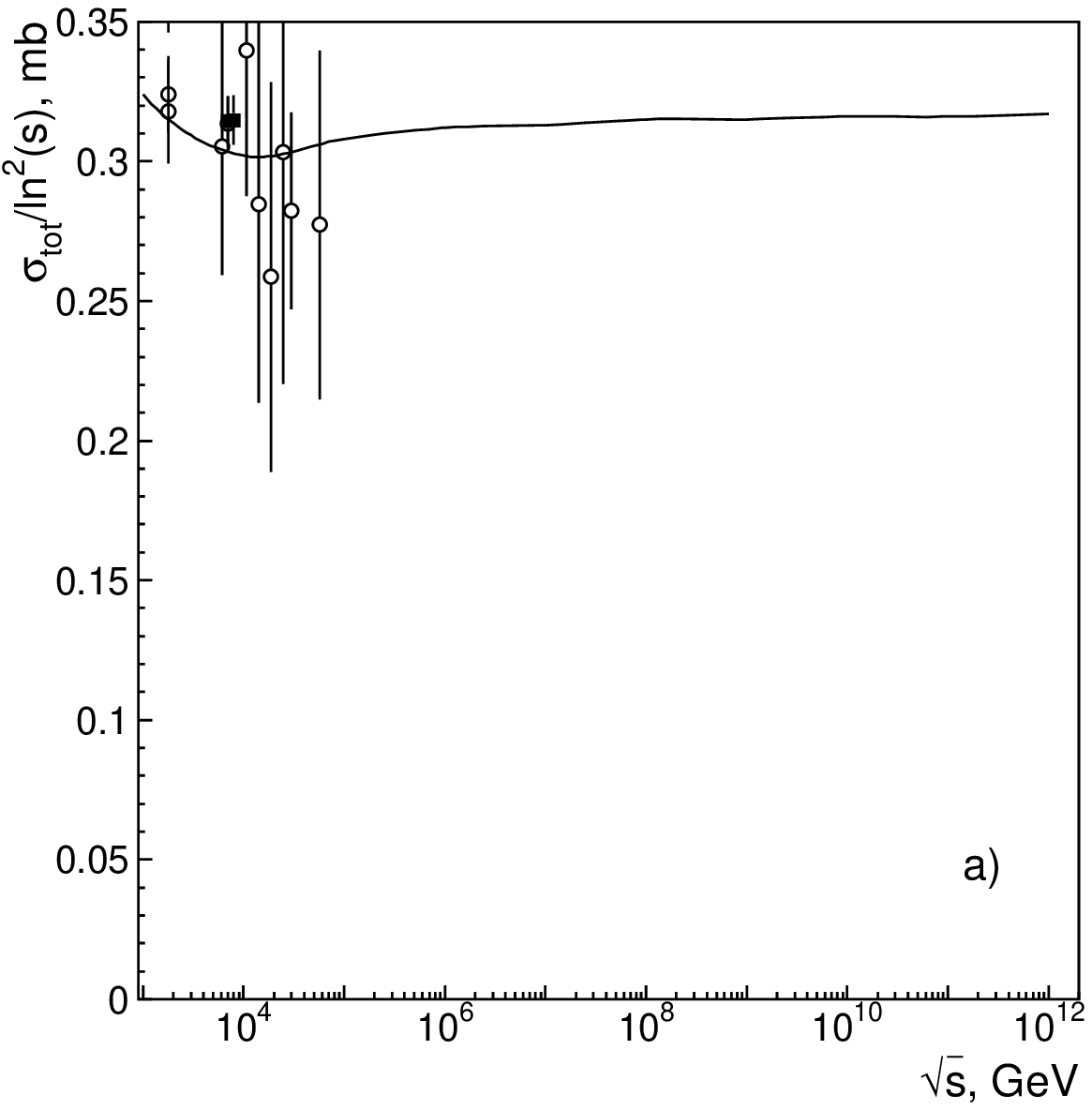,width=8cm}}
\centerline{ \epsfig{file=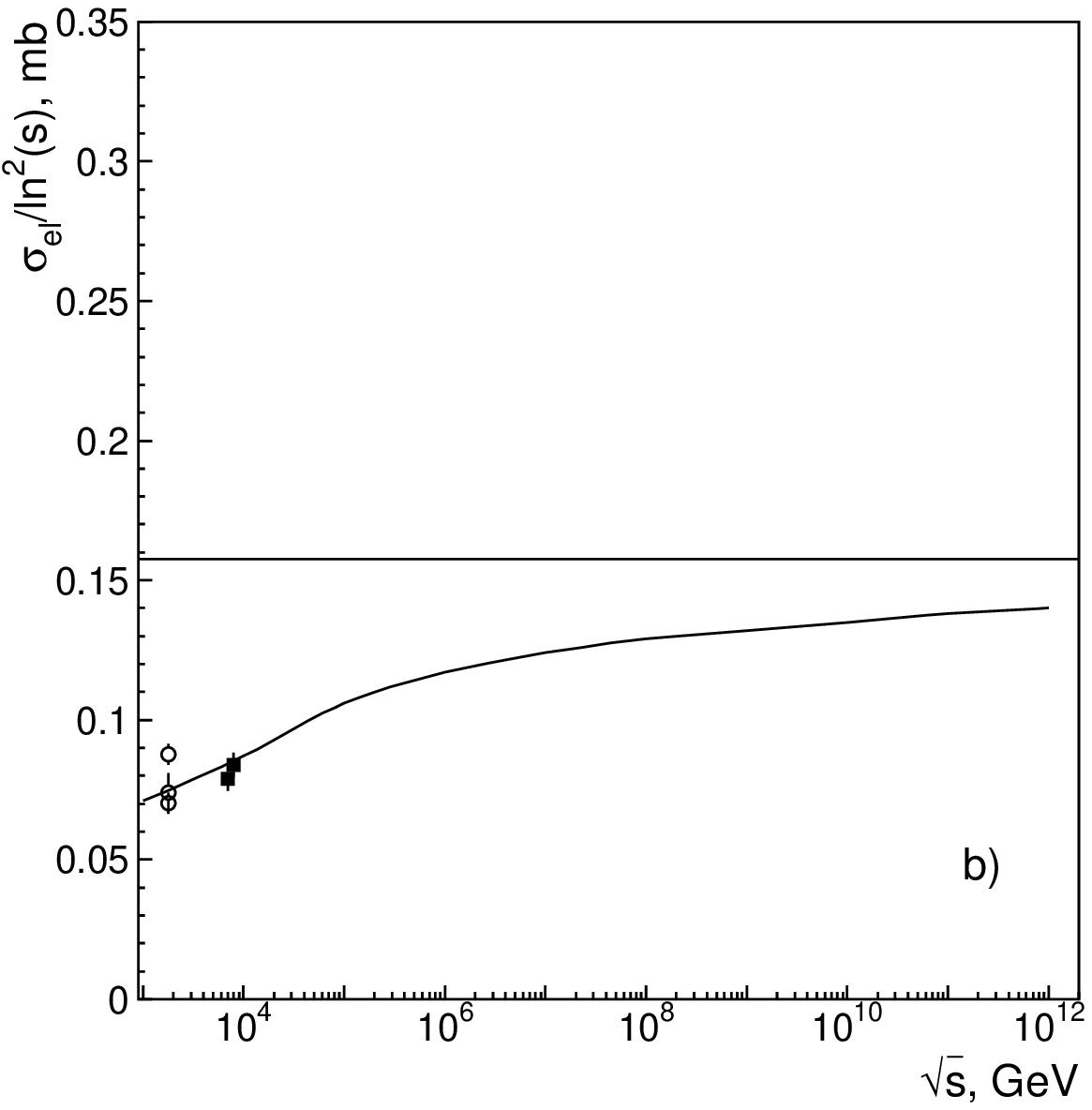,width=8cm}}
\caption{Total and elastic cross section data \cite{pre,totem,auger}
versus fit curves for (a) $\sigma_{tot}(pp)/\ln^2s$ and
(b) $\sigma_{el}(pp)/\ln^2s$, where $\sqrt s$ in GeV units, in the
Dakhno-Nikonov model for
the energy region $\sqrt{s}>1$ TeV.
The straight line is the asymptotic limit for the elastic cross section:
$\sigma_{el}(s)=\sigma^{(asym)}_{tot}/2$.
}
\label{f2}
\end{figure}

\begin{figure}[h]
\centerline{\epsfig{file=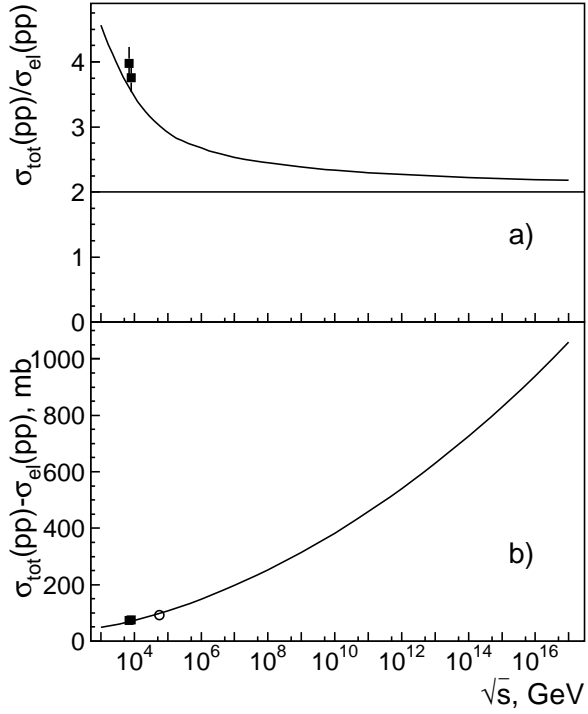,width=8cm}}
\caption{
Proton-proton collisions:
a) the ratio $\sigma_{tot}(pp)/\sigma_{el}(p p)\to 2$ and
b) its difference,
$\sigma_{inel}(p p)=\sigma_{tot}(pp)-\sigma_{el}(p p)\propto \ln^2 s$; data
from refs. \cite{totem,auger}.
 \label{pi-p-1}  }
 \end{figure}

\begin{figure}[ht]
\centerline{\epsfig{file=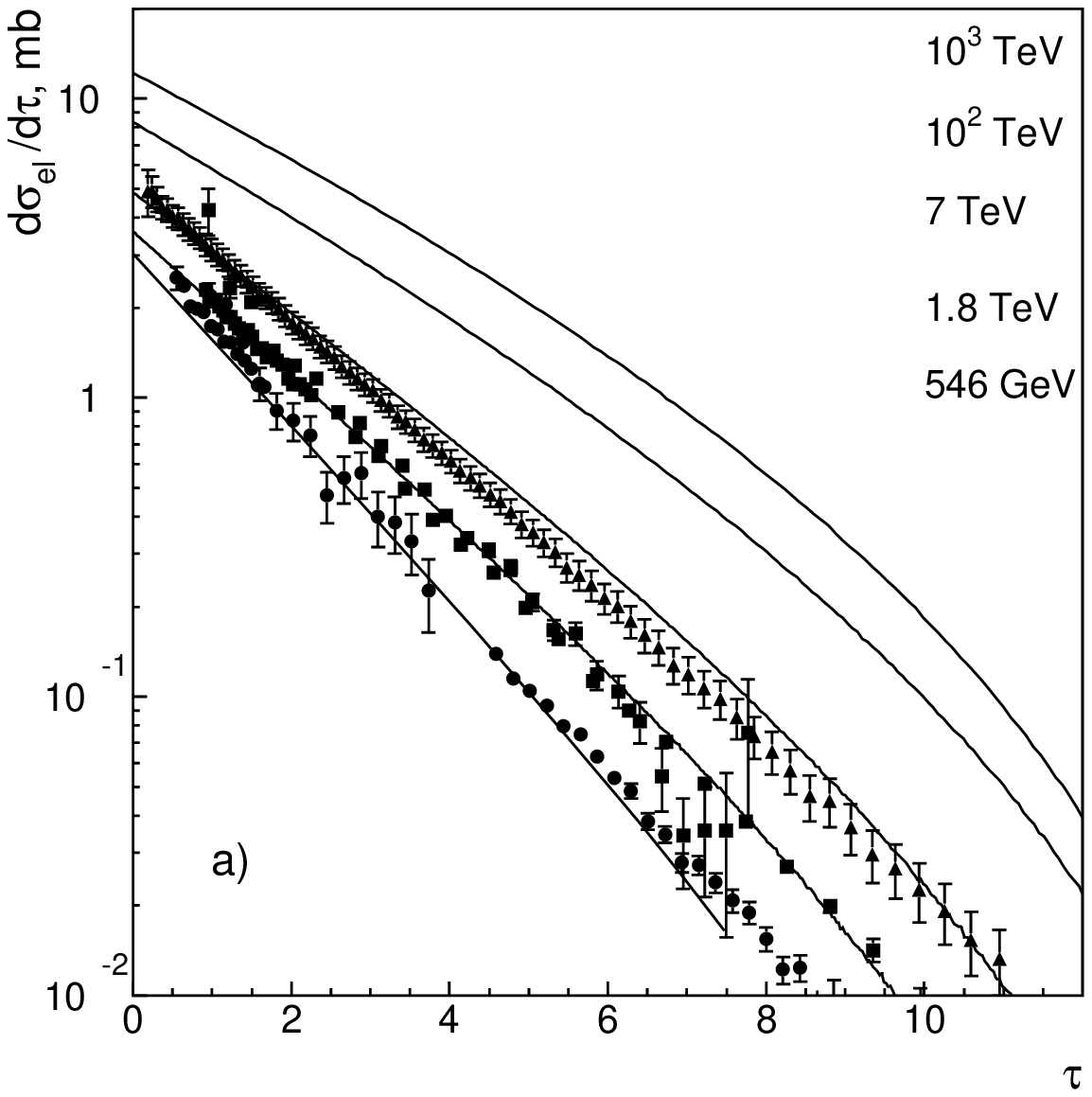,width=8cm}}
\centerline{\epsfig{file=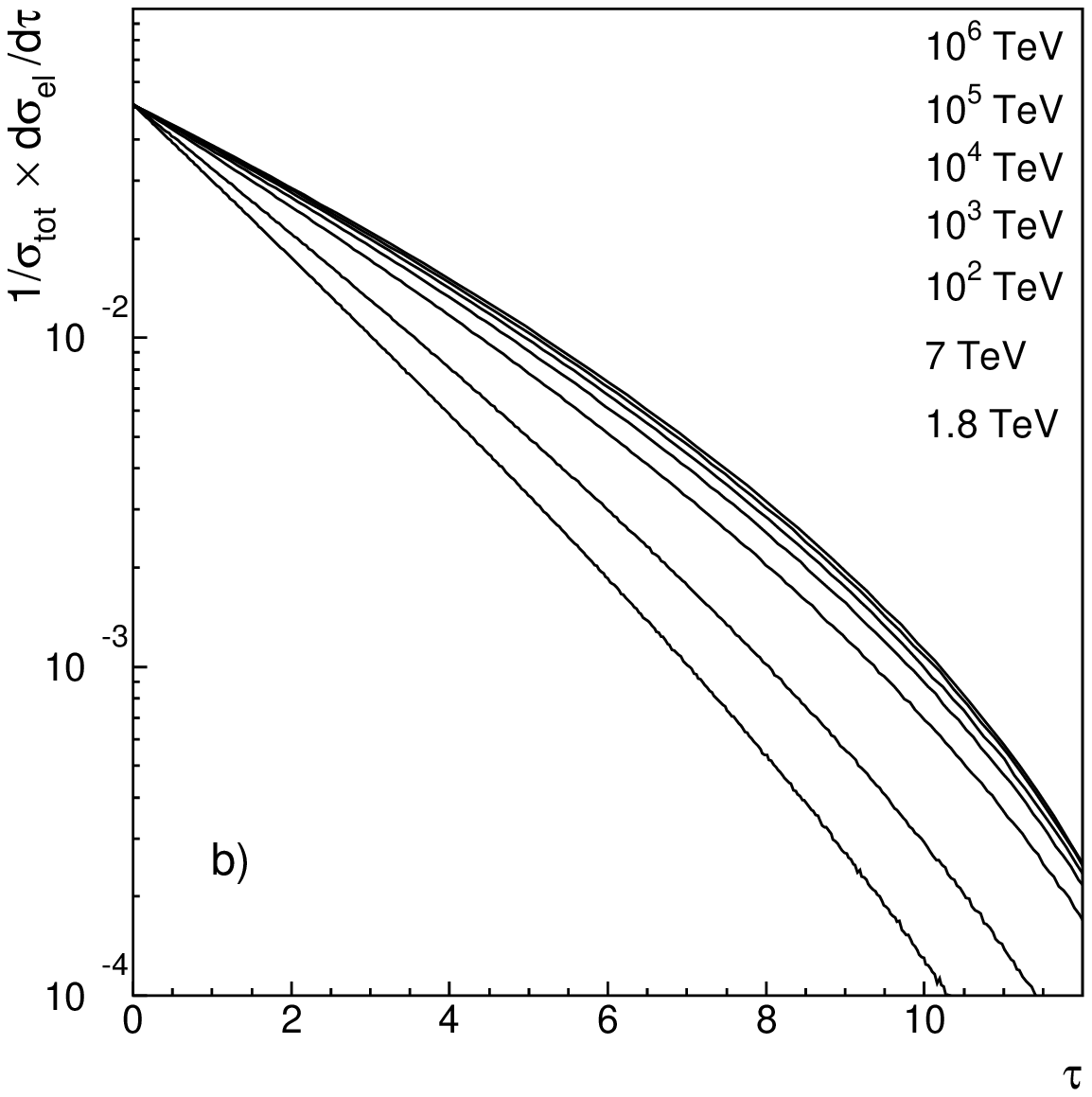,width=8cm}}
\caption{ a) Differential cross sections
$ d\sigma_{el}/d\tau$, where $\tau=\sigma_{tot}{\bf q}^2_\perp$,
  at $\sqrt s =0.546,\, 1.8,\, 7.0$
 TeV and their descriptions in the Dakhno-Nikonov model;
 b)  Calculated differential cross sections
$1/\sigma_{tot}\times d\sigma_{el}/d\tau$ at
$\sqrt{ s} = 1,\,10,\, 100,\, 1000,...10^6$ TeV
and their approach
to the $\tau$-scaling limit.
}
\label{f4}
\end{figure}

\section{Black disk picture and predictions for
the ultrahigh energy region, $\sqrt{s} > 10^2$ TeV }

The predictions we give are definitely related to the picture of the
black disk. The freedom of predictions is connected to the rate of
the black disk radius growth and the detailed structure of the black
disk boundary. Just to fix this freedom we use the Dakhno-Nikonov
model considering it as reasonably realistic. The
parameters are
\be
\label{yf14}
\begin{tabular}{l|r}
parameters  &   ref.\, \cite{ann} \\
\hline
$\Delta$                     &  0.273 \\
$g^2_0$ [mb]                 &  8.106 \\
$g^2_1$ [mb/GeV$^{2\Delta}$] &  0.379 \\
$\alpha'_P$ [(GeV/c)$^{-2}$] &  0.129 \\
$G$ [(GeV/c)$^{-2}$]         & -0.365 \\
$r^2_{cs}$ [(GeV/c)$^{-2}$]  &  0.670 \\
\end{tabular}
\ee
The profile function $T(b)$, calculated using the parameters of
Eq. (\ref{yf14}), predicts  the black disk regime at
$\sqrt{s}\ga 10^2$ TeV.
The profile function is determined as follows:
\bea
\sigma_{tot}&=&2\int d^2b\; T(b)=2\int d^2b
\Big[1-e^{-\frac12\chi(b)}\Big], \nn \\
4\pi\frac{d\sigma_{el}}{d{\bf q}^2_\perp}&=& A^2({\bf
q}^2_\perp),
\nn \\
A({\bf q}_\perp)&=&\int d^2b e^{i{\bf b}{\bf
q}_\perp} T(b).
\label{yf13}
\eea
This is shown in Fig. \ref{f5}a
for pre-LHC, LHC and ultrahigh energies. The profile functions
$T(b)$ are not saturated at $\sqrt{s}\sim 0.5-2.0$ TeV being $T(b)<
1$. According to the fit, a black spot ($T(b)\simeq 1$ at $b<
2$GeV$^{-1}\simeq 0.4fm$) appears at $\sqrt{s}\sim 50-100$ TeV; this
phenomenon indicates the start of the black disk regime.
 At $\ln s>>1$, when the asymptotic regime works, there are two
  clear regions in the $b$-space (Fig. \ref{f5}a):
 with $T(b)\simeq 1$ (black disk area) and $T(b)\simeq 0$ (transparent
 area).
 Conventionally we determine these areas by the constraints
\bea  \label{yf15}
&&
{\bf b}^2 < 4\Delta\alpha'_P\ln^2{\frac {s}{s_-}}\,,\qquad
{\rm with}\quad T(b)>0.97,
\nn \\
&& {\bf b}^2 > 4\Delta\alpha'_P\ln^2{\frac {s}{s_+}}\,,\qquad {\rm
with}\quad T(b)< 0.03\,,
\quad
\eea
 giving the black disk radius:
\be \label{yf16}
R_{black}=2\sqrt{ \Delta\alpha'_P}\,\ln{\frac
{s}{s_R}}\,,\qquad \sqrt{s_{R}}\simeq 80\, {\rm GeV}\,.
\ee
 The black disk radius depends on parameters of the leading pomeron
only, (factor $\sqrt{\Delta\alpha'_P}$), which results in Gribov's
universality of hadronic total cross sections at asymptotic energies
\cite{Gribov-tot}.

The growth of the profile function, $\delta T(b, \sqrt{s})\equiv
[T(b,10 \sqrt{s})-T(b, \sqrt{s})]$, is  demonstrated in Fig.
\ref{f5}b. Steady values of areas under $\delta T(b, \sqrt{s})$ tell
us that the ratio of contributions of the border region to that of
the internal disk region is decreasing with energy
$\sigma_{inel}^{border}/\sigma_{inel}^{intern} \to 1/\ln s$ . It is
revealed as a basis for the black disk description of asymptotic
cross sections.

In Fig. \ref{f2} we show total and elastic cross sections in the
$\sqrt{s}\sim 1- 100$ TeV region \cite{pre,totem,auger} and their
description in the fit of ref. \cite{ann}. An extension of the
fitting curves into the $\sqrt{s}>100$ TeV region tells us that we
have a relatively fast approach  to the asymptotic behaviour for
$\sigma_{tot}/\ln ^2 s$; the approach of $\sigma_{el}/\ln ^2 s$ to
the asymptotic value is slow. A slow switching on of asymptotics is
definitely seen in the ratio $\sigma_{tot}(pp)/\sigma_{el}(p p)$,
Fig. \ref{pi-p-1}a. The inelastic cross section $\sigma_{inel}(p
p)=\sigma_{tot}(pp)-\sigma_{el}(p p)\propto \ln^2 s$ is demonstrated
in Fig. \ref{pi-p-1}b.

In Fig. \ref{f4}a we show $d\sigma_{el}/d\tau$ (let us recall that
$\tau=\sigma_{tot} {\bf q}^2_\perp) $ at ISR and LHC energies; the
approach of $1/\sigma_{tot}\times d\sigma_{el}/d\tau$ to the
$\tau$-scaling limit is demonstrated in Fig. \ref{f4}b .

\begin{figure}[h]
\centerline{\epsfig{file=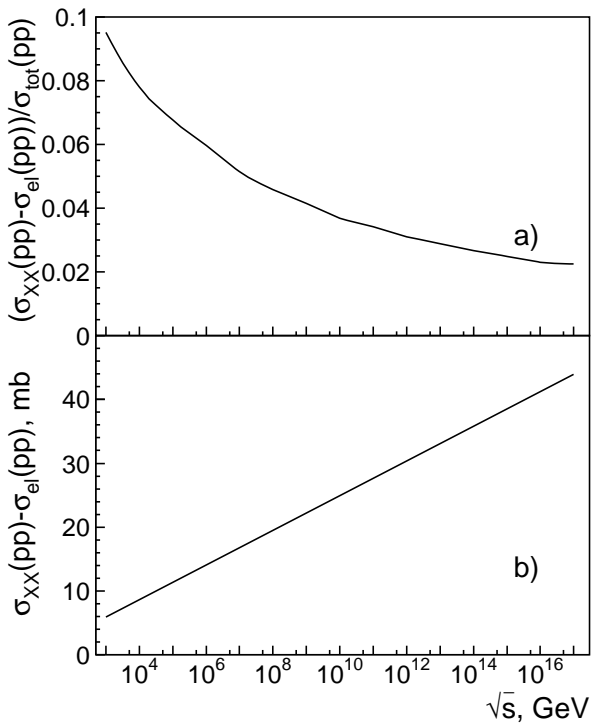,width=8cm}}
\caption{a) The ratio given in Eq. (\ref{ddpp})
and
b) the difference
$\sigma_{X(p)X(p)}(pp)-\sigma_{el}(pp)=2\sigma_{D(p)}(pp)+\sigma_{D(p)D(p)}(pp)\propto \ln s$.
 \label{dpp_02} }
 \end{figure}

In Fig. \ref{dpp_02}a we show  the ratio
\be \label{ddpp}
\frac{\sigma_{X(p)X(p)}(pp)-\sigma_{el}(pp)}{\sigma_{tot}(pp)}
=\frac{2\sigma_{D(p)}(pp)+\sigma_{D(p)D(p)}(pp)}{\sigma_{tot}(pp)},
 \ee
see Eq. (\ref{dd}). It tends to zero at $\ln s\to\infty$, this fact is
in complete agreement with
$\sigma_{tot}/\sigma_{el}\to 2$.
In Fig. \ref{dpp_02}b we show the sum
$2\sigma_{D(p)}(pp)+\sigma_{D(p)D(p)}(pp)$,
it increases as
\bea
\sigma_{X(p)X(p)}(pp)-\sigma_{el}(pp)\simeq 0.58
\ln\frac{s}{s^{XX}_{el}}\; {\rm mb},
\nn \\
s^{XX}_{el}\simeq 1.22 {\rm GeV }^2.
\eea

The ratio $Re \,A_{el}/Im\, A_{el}$ at ${\bf q}^2_\perp \simeq 0$ at
asymptotic energies is determined by the analyticity of the
scattering amplitude:
\be \label{yf17} A_{el}\propto i[\ln^2(s/s_0)+\ln^2 (-s/s_0)],
\quad \frac{Re\,
A_{el}}{Im\,A_{el}}\simeq \frac{\pi}{\ln(s/s_0)}
\ee
 This estimation
gives us $Re \,A_{el}/Im\, A_{el}=0.18\pm 0.04$ for the $\sqrt{s}\ga
100$ TeV region, being in qualitative agreement with the 7-TeV data:
$Re \,A_{el}/Im\, A_{el}=0.14^{+ 0.01}_{-0.08}$ \cite{totem} .

The extension of the results on $\pi p$ collisions
does not cause problems in the Dakhno-Nikonov model since for that
the addition of the pion quark distribution is needed only. The
distribution of quarks in a pion is known (see, for example, ref.
\cite{AMN}) and this allows to give predictions.

\begin{figure}[h]
\centerline{\epsfig{file=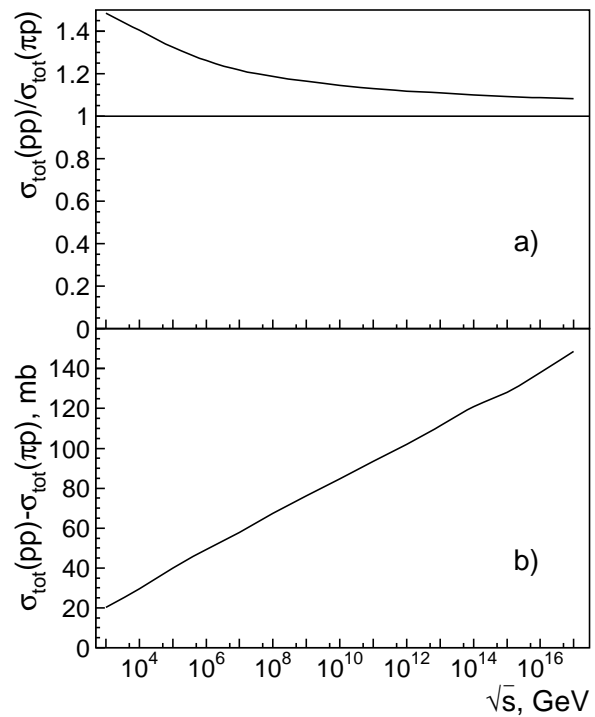,width=8cm}}
\caption{a) Ratio of proton-proton and pion-proton total cross
sections, $\sigma_{tot}(pp)/\sigma_{tot}(\pi p)\to 1$, and
b) its difference $\sigma_{tot}(pp)-\sigma_{tot}(\pi p)\propto \ln s$.
 \label{pi-p-03} }
 \end{figure}

\begin{figure}[ht]
\centerline{\epsfig{file=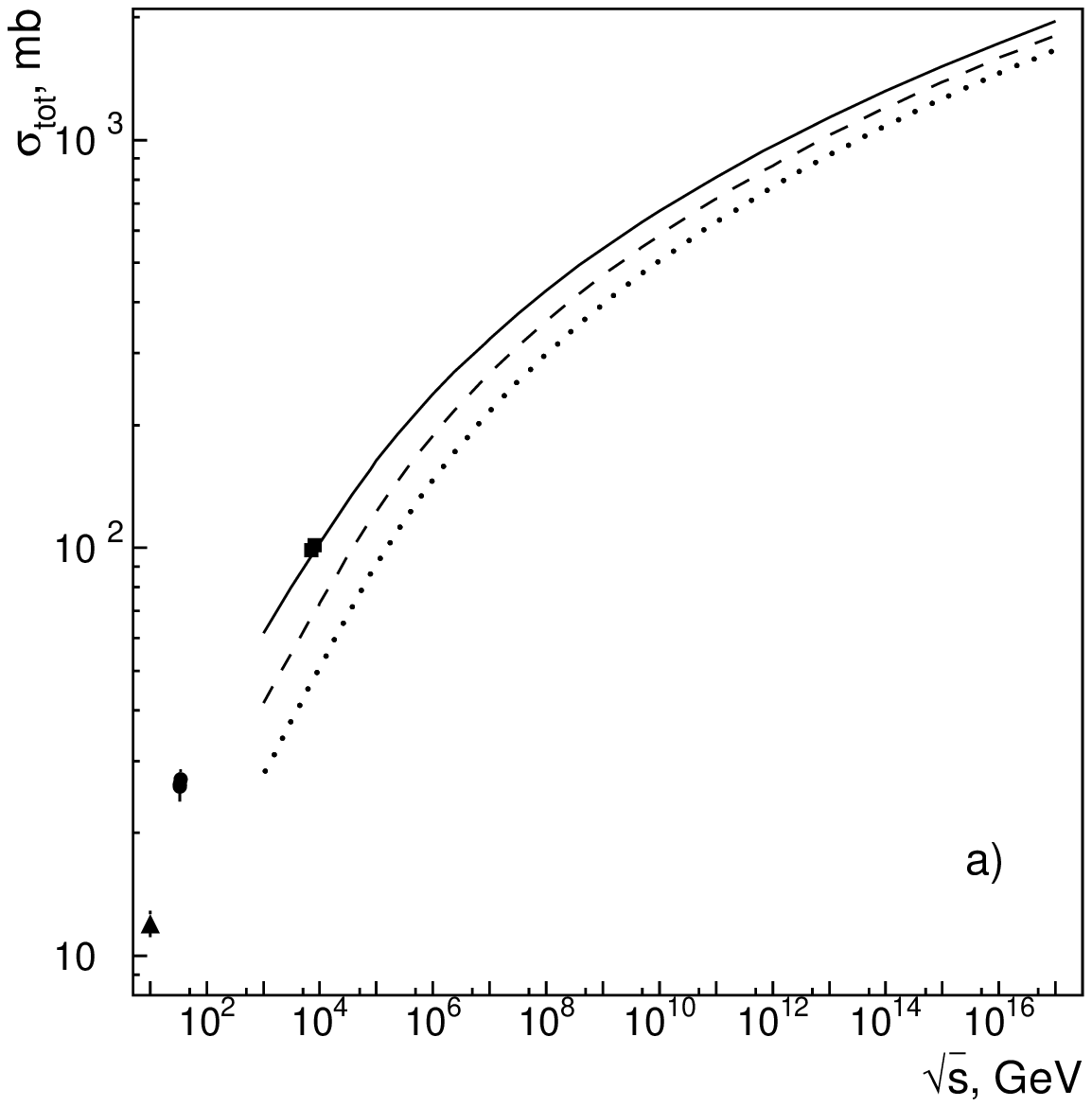,width=8cm}}
\centerline{\epsfig{file=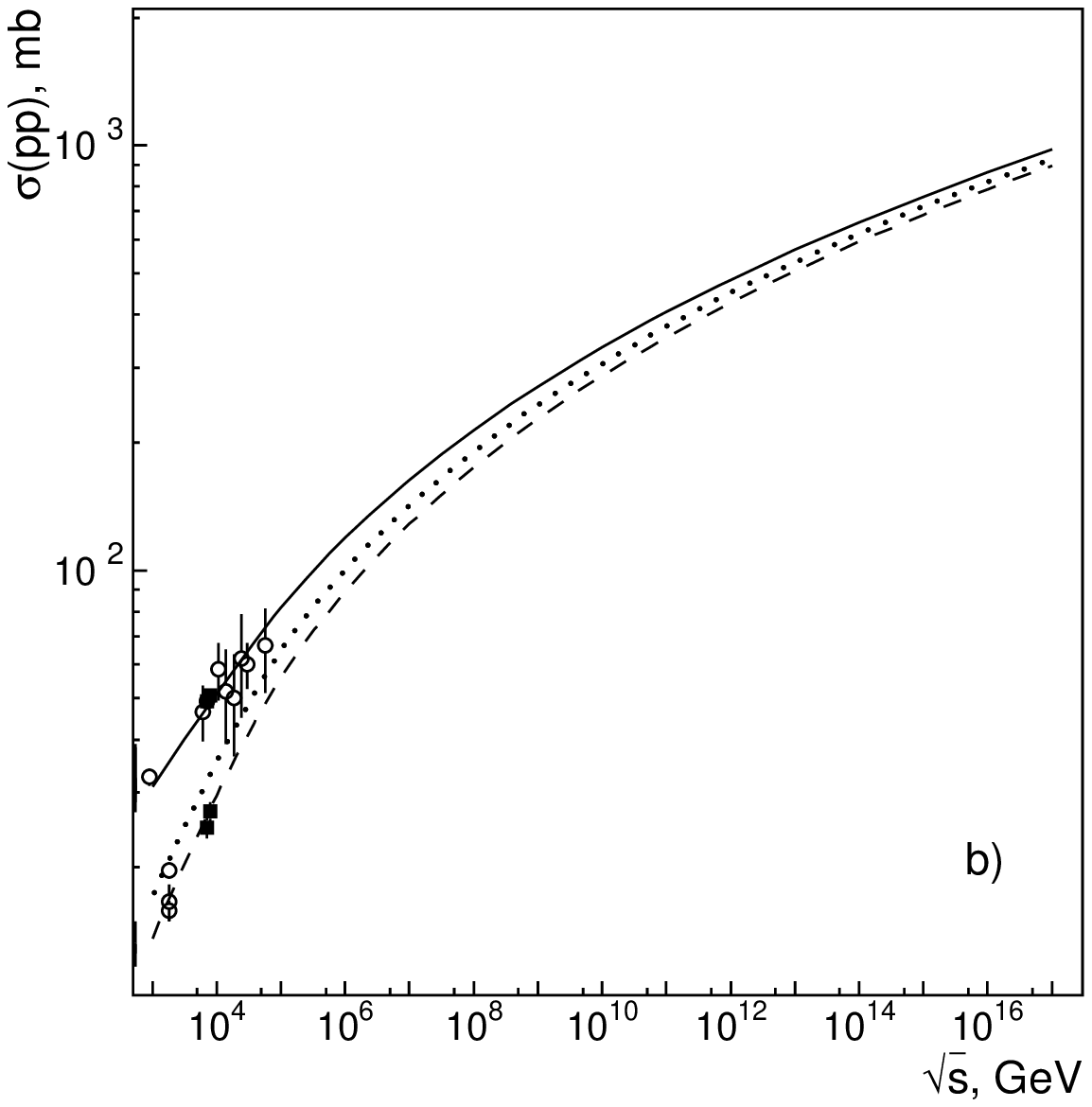,width=8cm}}
\caption{ a) Total cross sections: $\sigma_{tot}(pp)$ (solid line),
$\sigma_{tot}(\pi p)$ (dashed line) and
$\sigma_{tot}(\pi \pi)$ (dotted line).
 Squares ($pp$) are from
\cite{totem},
circles ($\pi p$) are from \cite{Dersch:1999zg},
triangles ($\pi \pi$) are from \cite{Abramowicz:1979ca}.
b) Proton-proton
collisions: $\frac12\sigma_{tot}(pp)$ (solid line), $\sigma_{el}(pp)
+2\sigma_{D(p)}(pp)$ (dotted line) and $\sigma_{el}(pp)$ (dashed line).
Data \cite{pre,totem,auger} stand for $\frac12\sigma_{tot}(pp)$ and
$\sigma_{el}(pp) $.
 }
\label{f6}
\end{figure}

The ratio $\sigma_{tot}(pp)/\sigma_{tot}(\pi p)$ is shown in Fig.
\ref{pi-p-03}a; it asymptotically tends to 1. The difference of
proton-proton and pion-proton total cross sections    increases
with energy and can be described at $\sqrt{s}\ga 10^6$ GeV as
\be
\sigma_{tot}(pp)-\sigma_{tot}(\pi p)\simeq 1.91
\ln{\frac{s}{s^{pp}_{\pi p}}}\; {\rm mb},\quad
s^{pp}_{\pi p}\simeq 6.25 {\rm GeV }^2.
 \label{pi-p-dif}
 \ee
 The universality of the total cross sections means the equality of the
leading terms ($\propto \ln^2 s$) only, and it is demonstrated in
Fig. \ref{f6}a where $\sigma_{tot}(pp)$, $\sigma_{tot}(\pi p)$ and
$\sigma_{tot}(\pi \pi)$ are shown. For comparison we show the
approach to the asymptotic limit of diffractive cross sections in
$pp$ collisions, Fig. \ref{f6}b : the Miettinen-Pumplin limit value
$\frac 12 \sigma_{tot}(pp)$, elastic plus single diffraction
dissociation processes $\sigma_{el}(pp)+2\sigma_{D(p)}(pp)$ and
$\sigma_{el}(pp)$. Here we definitely see a more rapid switching on
of the asymptotic regime.

We perform the unitarization of the scattering amplitude supposing it
originates from
conventional pomerons, though other types of     input
pomerons are possible as well as non-pomeron short-range
contributions (for example, see \cite{KL,KN,GLM,AKL}). But
here we concentrate our attention on
peripheral interactions and their
transformation with     energy growth. Small deviations of the fitting
curves from data can be easily improved using
some kind of short-range contributions.

      \section{Conclusion}

In ref. \cite{ann} the description of
diffractive $pp$ collisions was performed in terms of
the Dakhno-Nikonov model for pre-LHC and
LHC  energies: with parameters found in \cite{ann} we study here the
ultrahigh energy region where the asymptotic regime works.

The twofold structure of hadrons -- hadrons are built from constituent
quarks; the latter are formed by clouds of partons which manifest
itself in high energy hadron collisions. At moderately high energies
colliding protons reveal themselves in     impact parameter space as
three disks corresponding to three constituent quarks.
At ultrahigh energies the situation is transformed to a one-disk
picture,
and the energy of this transformation is that of LHC. The radius of
the black disk at asymptotic energies    increases  as $\ln{s}$,
hence providing a $\ln^2s$ growth of  $\sigma_{tot}$, $\sigma_{el}$
with $\sigma_{el}/\sigma_{tot}\to 1/2$ and a $\tau$-scaling for
diffractive cross sections.

The calculations we have carried out demonstrate a comparatively
fast approach of $\sigma_{tot}(s)$ to     asymptotic behaviour (Fig.
\ref{f2}a),      in contrast to $\sigma_{el}(s)$ (Fig. \ref{f2}b).
It also means a slow approach of
$\sigma_{inel}(s)=\sigma_{tot}(s)-\sigma_{el}(s)$ to the asymptotic
behaviour.

A good level of description of the diffractive $pp$ scattering is
demonstrated on Fig. \ref{f4}. It is seen that the $\tau$-scaling
for $1/\sigma_{tot}\times d\sigma_{el}/d\tau$ is at work within
10$\%$ accuracy at $\sqrt s \ga 100$ TeV (the upper curves in
Fig. \ref{f4}b).

The change of the regime, from the constituent quark collision
picture to that with a united single disk, was discussed
\cite{ani-she,ani-lev-rys,ani-bra-sha} when definite indications
about hadron cross section growth appeared. It  is emphasized that
the single black disk regime should change probabilities of
productions of hadrons in the fragmentation region (hadrons with
$x=p/p_{in}\sim 1$); for a more detailed discussion of the scaling
violation in the hadron fragmentation region see \cite{book3}.


 Diffractive dissociation processes are
increasing at asymptotic energies. For $pp$ collisions
the relative weights
of quasi-elastic cross sections are vanishing with the energy
growth,
$\sigma_{D(p)}(pp\to pX)/\sigma_{tot}(pp)\to 0$ and
$\sigma_{D(p)D(p)}(pp\to XX)/\sigma_{tot}(pp)\to 0$,
while the cross sections
increases,
$2\sigma_{D(p)}(pp\to pX)+\sigma_{D(p)D(p)}(pp\to XX)\simeq
0.58\ln{s/s^{XX}_{el}}$ mb.
It means we can estimate diffractive production cross section  of
$N_{\frac 12^+}(1440)$ as
$\left(\frac 12\div\frac{1}{10}\right)\;0.58\ln{s/s^{XX}_{el}}$ mb.


A steady increase of the black disk radius $R_{black}\propto \sqrt{
\Delta\alpha'_P}\ln s$,
is determined by parameters of the leading $t$-channel singularity:
the pomeron intercept $\alpha(0)=1+\Delta$ ($\Delta>0$) and the
pomeron trajectory slope $\alpha'_P$. The $s$-channel unitarization
of the scattering amplitude damps the strong pomeron pole
singularity, transforming it into a multipomeron one. Therefore, we
face  an intersection of problems of the gluon content of
$t$-channel states at ultrahigh energies and the physics of gluonic
states, glueballs. At present the glueball states are subjects of
discussions,              see, for example
\cite{vva-usp,klempt-zaitsev,ochs} and references therein.  Studies
of phenomenons related to glueballs and multigluon states at small
$|t|$ (or, at small masses) are enlightening for the confinement
singularity - see discussion in ref. \cite{conf-PR}. The large value
of mass of the soft effective gluon (and the corresponding value the
low-lying glueballs) and the slow rate of the black disk increase
appear to be related phenomena.

\section{Acknowledgment}

We thank Ya.I. Azimov, D.V. Bugg, A.K. Likhoded,  M.G. Ryskin,
and A.V. Sarantsev for useful discussions and comments.
 The work was supported by grants RFBR-13-02-00425 and
 RSGSS-4801.2012.2 .

   \end{document}